\documentstyle[11pt,epsfig,psfig,amsmath]{article}
\begin{document}

\title{{\bf Surface vacuum energy and stresses on a plate uniformly
accelerated through the Fulling-Rindler vacuum}}
\author{ A. A. Saharian$^{1}$\thanks{%
E-mail: saharyan@www.physdep.r.am } and M. R. Setare$^{2}$\thanks{%
E-mail: rezakord@ipm.ir} \\
{\it $^1$ Department of Physics, Yerevan State University, Yerevan, Armenia }
\\
{\it $^2$ Institute for Theoretical Physics and Mathematics, Tehran, Iran}}
\maketitle

\begin{abstract}
The vacuum expectation value of the surface energy-momentum tensor is
evaluated for a massless scalar field obeying mixed boundary condition on an
infinite plate moving by uniform proper acceleration through the
Fulling-Rindler vacuum. The generalized zeta function technique is used, in
combination with the contour integral representation. The surface energies
for separate regions on the left and on the right of the plate contain pole
and finite contributions. Analytic expressions for both these contributions
are derived. For a minimally coupled scalar the surface energy-momentum
tensor induced by vacuum quantum effects corresponds to a source of the
cosmological constant type located on the plate and with the cosmological
constant determined by the proper acceleration of the plate.
\end{abstract}

\bigskip

PACS number(s): 03.70.+k, 11.10.Kk

\newpage

\section{Introduction}

\label{sec:Int}

The use of general coordinate transformations in quantum field
theory leads to an infinite number of unitary inequivalent
representations of the commutation relations. Different
inequivalent representations will in general give rise to
different pictures with different physical implications, in
particular to different vacuum states. For instance, the vacuum
state for a uniformly accelerated observer, the Fulling--Rindler
vacuum \cite{Full73,Unru76,Boul75,Gerl89}, turns out to be
inequivalent to that for an inertial observer, the familiar
Minkowski vacuum. Quantum field theory in accelerated systems
contains many special features produced by a gravitational field.
This fact allows one to avoid some of the difficulties entailed by
renormalization in a curved spacetime. In particular, the near
horizon geometry of most black holes is well approximated by the
Rindler metric and a better understanding of physical effects in
this background could serve as a handle to deal with more
complicated geometries like Schwarzschild. The Rindler geometry
shares most of the qualitative features of black holes and is
simple enough to allow detailed analysis. Another motivation for
the investigation of quantum effects in the Rindler space is
related to the fact that this space is conformally related to the
de Sitter (dS) space and to the Robertson--Walker space with
negative spatial curvature. As a result the expectation values of
the energy--momentum tensor for a conformally invariant field and
for corresponding conformally transformed boundaries on dS and
Robertson--Walker backgrounds can be derived from the
corresponding Rindler counterpart by the standard transformation
(see, for instance, \cite{Birrell}).

The problem of vacuum polarization by an infinite plane boundary moving with
uniform acceleration through the Fulling-Rindler vacuum was investigated by
Candelas and Deutsch \cite{Candelas} for the conformally coupled $4D$
Dirichlet and Neumann massless scalar and electromagnetic fields. In this
paper only the region of the right Rindler wedge to the right of the barrier
is considered. In Ref. \cite{Saha02} we have investigated the Wightman
function and the vacuum expectation values of the energy momentum-tensor for
the massive scalar field with general curvature coupling parameter,
satisfying the Robin boundary conditions on the infinite plane in an
arbitrary number of spacetime dimensions and for the electromagnetic field.
Unlike Ref. \cite{Candelas} we have considered both regions, including the
one between the barrier and Rindler horizon. The vacuum expectation values
of the energy-momentum tensors for Dirichlet or Neumann scalar and
electromagnetic fields for the geometry of two parallel plates moving by
uniform acceleration are investigated in Ref. \cite{Avag02}. In particular,
the vacuum forces acting on the boundaries are evaluated. They are presented
as a sum of the interaction and self-action parts. The interaction forces
between the plates are always attractive for both scalar and electromagnetic
cases. In Refs. \cite{Saha02,Avag02} the mode summation method is used in
combination with the generalized Abel-Plana summation formula \cite{Sahrev}.
This allowed us to present the vacuum expectation values in terms of the
purely Rindler and boundary-induced parts. Due to the well-known
non-integrable surface divergences in the boundary parts, the total Casimir
energy cannot be obtained by direct integration of the vacuum energy density
and needs an additional regularization. In Ref. \cite{Saha04a} the Casimir
energy is evaluated for massless scalar fields under Dirichlet or Neumann
boundary conditions, and for the electromagnetic field with perfect
conductor boundary conditions on one and two infinite parallel plates moving
by uniform proper acceleration through the Fulling--Rindler vacuum in an
arbitrary number of spacetime dimension. In Ref. \cite{Saha04c} the
conformal relation between dS and Rindler spacetimes and the results from
Ref. \cite{Saha02} are used to generate the vacuum expectation values of the
energy-momentum tensor for a conformally coupled scalar field in dS
spacetime in presence of a curved brane on which the field obeys the Robin
boundary condition with coordinate dependent coefficients.

For scalar fields with general curvature coupling in Ref. \cite{Rome02} it
has been shown that in the discussion of the relation between the mode sum
energy, evaluated as the sum of the zero-point energies for each normal mode
of frequency, and the volume integral of the renormalized energy density for
the Robin parallel plates geometry it is necessary to include in the energy
a surface term concentrated on the boundary (see also the discussion in Ref.
\cite{Full03}). Similar issues for the spherical and cylindrical boundary
geometries are discussed in Refs. \cite{Saha01,Rome01}. An expression of the
surface energy-momentum tensor for a scalar field with general curvature
coupling parameter in the general case of bulk and boundary geometries is
derived in Ref. \cite{Saha04b}. In the present paper, by using this
expression and the zeta function technique, we evaluate the vacuum
expectation value of the surface energy-momentum tensor for a plate moving
by uniform proper acceleration through the Fulling-Rindler vacuum.

The paper is organized as follows. In Section \ref{sec:RRreg} the surface
energy density and vacuum stresses are evaluated for the region on the right
of the plate (RR region). We construct an integral representation for the
related zeta function and analytically continue it to the physical region.
Similar problem for the region on the left of the plate (RL region) is
investigated in Section \ref{sec:RLreg}. Section \ref{sec:Conc} concludes
the main results of the paper.

\section{Surface energy-momentum tensor in the RR region}

\label{sec:RRreg}

Consider a real massless scalar field $\varphi (x)$ with curvature coupling
parameter $\zeta $ satisfying the field equation
\begin{equation}
\nabla _{l}\nabla ^{l}\varphi +\zeta R\varphi =0,  \label{fieldeq}
\end{equation}%
with $R$ being the scalar curvature for a $(D+1)$-dimensional background
spacetime, $\nabla _{l}$ is the covariant derivative operator associated
with the corresponding metric tensor $g_{ik}$. For minimally and conformally
coupled scalars one has $\zeta =0$ and $\zeta =(D-1)/4D$, respectively. Our
main interest in this paper will be the surface Casimir energy and stresses
induced on a single plate moving with uniform proper acceleration when the
quantum field is prepared in the Fulling-Rindler vacuum. We will assume that
the field satisfies the mixed boundary condition
\begin{equation}
(A_{s}+n^{l}\nabla _{l})\varphi (x)=0  \label{boundcond}
\end{equation}%
on the plate, where $A_{s}$ is a constant, $n^{l}$ is the unit inward normal
to the plate. For this boundary condition the vacuum expectation values of
the bulk energy-momentum tensor in the both RR and RL regions are evaluated
in Ref. \cite{Saha02}. In Ref. \cite{Saha04b} it was argued that the
energy-momentum tensor for a scalar field on manifolds with boundaries in
addition to the bulk part contains a contribution located on the boundary.
For the boundary $\partial M_{s}$ the surface part of the energy-momentum
tensor is presented in the form \cite{Saha04b}
\begin{equation}
T_{ik}^{{\rm (surf)}}=\delta (x;\partial M_{s})\tau _{ik}  \label{Ttausurf}
\end{equation}%
with
\begin{equation}
\tau _{ik}=\zeta \varphi ^{2}K_{ik}-(2\zeta -1/2)h_{ik}\varphi n^{l}\nabla
_{l}\varphi ,  \label{tausurf}
\end{equation}%
and the "one-sided" delta-function $\delta (x;\partial M_{s})$ locates this
tensor on $\partial M_{s}$. In Eq. (\ref{tausurf}), $K_{ik}$ is the
extrinsic curvature tensor of the boundary $\partial M_{s}$ and $h_{ik}$ is
the corresponding induced metric. Let $\{\varphi _{\alpha }(x),\varphi
_{\alpha }^{\ast }(x)\}$ be a complete set of positive and negative
frequency solutions to the field equation (\ref{fieldeq}), obeying boundary
condition (\ref{boundcond}). Here $\alpha $ denotes a set of quantum numbers
specifying the solution. By expanding the field operator over the
eigenfunctions $\varphi _{\alpha }(x)$, using the standard commutation rules
and the definition of the vacuum state, for the vacuum expectation value of
the surface energy-momentum tensor one finds
\begin{equation}
\langle 0|T_{ik}^{{\rm (surf)}}|0\rangle =\delta (x;\partial M_{s})\langle
0|\tau _{ik}|0\rangle ,\quad \langle 0|\tau _{ik}|0\rangle =\sum_{\alpha
}\tau _{ik}\{\varphi _{\alpha }(x),\varphi _{\alpha }^{\ast }(x)\},
\label{modesumform}
\end{equation}%
where $|0\rangle $ is the amplitude for the corresponding vacuum state, and
the bilinear form $\tau _{ik}\{\varphi ,\psi \}$ on the right of the second
formula is determined by the classical energy-momentum tensor (\ref{tausurf}%
).

In the accelerated frame it is convenient to introduce Rindler coordinates $%
(\tau ,\xi ,{\bf x})$ which are related to the Minkowski ones, $(t,x^{1},%
{\bf x})$ by transformations
\begin{equation}
t=\xi \sinh \tau ,\quad x^{1}=\xi \cosh \tau ,  \label{RindMin}
\end{equation}%
where ${\bf x}=(x^{2},\ldots ,x^{D})$ denotes the set of coordinates
parallel to the plate. In these coordinates one has $g_{00}=\xi ^{2}$, $%
g_{ik}=-\delta _{ik}$, $i\neq 0$, and a wordline defined by $\xi ,{\bf x}=%
{\rm const}$ describes an observer with constant proper acceleration $\xi
^{-1}$. Rindler time coordinate $\tau $ is proportional to the proper time
along a family of uniformly accelerated trajectories which fill the Rindler
wedge, with the proportionality constant equal to the acceleration. For the
geometry under consideration the metric and boundary conditions are static
and translational invariant in the hyperplane parallel to the plate. It
follows from here that the corresponding part of the eigenfunctions can be
taken in the standard plane wave form:
\begin{equation}
\varphi _{\alpha }=C_{D}\phi (\xi )\exp \left[ i\left( {\bf kx}-\omega \tau
\right) \right] ,\quad \alpha =({\bf k},\omega ),\quad {\bf k}=(k_{2},\ldots
,k_{D}).  \label{wavesracture}
\end{equation}%
The equation for $\phi (\xi )$ is obtained from field equation (\ref{fieldeq}%
). The corresponding linearly independent solutions are the Bessel modified
functions $I_{i\omega }(k\xi )$ and $K_{i\omega }(k\xi )$ of the imaginary
order, where $k=|{\bf k}|$. The eigenfrequencies are determined from the
boundary condition imposed on the field on the bounding surface. As such a
surface we take a plane boundary with coordinate $\xi =a>0$ corresponding to
a plate uniformly accelerated normal to itself with the proper acceleration $%
a^{-1}$. The plate divides the right Rindler wedge into two regions with $%
\xi >a$ (RR region) and $0<\xi <a$ (RL region). The vacuum properties in
these regions are different and we consider them separately.

For the RR region the unit normal to the boundary and nonzero components of
the extrinsic curvature tensor have the form
\begin{equation}
n^{l}=\delta _{1}^{l},\quad K_{00}=\xi ,  \label{normvec}
\end{equation}%
and $\phi (\xi )=K_{i\omega }(k\xi )$. For a given $ka$, the corresponding
eigenfrequencies $\omega =\omega _{n}=\omega _{n}(ka)$, $n=1,2,\ldots $, are
determined from boundary condition (\ref{boundcond}) and are solutions to
the equation
\begin{equation}
AK_{i\omega }(x)+xK_{i\omega }^{\prime }(x)=0,\quad x=ka,\quad A=A_{s}a,
\label{modeq}
\end{equation}%
where the prime denotes the differentiation with respect to the argument of
the function. For $A_{s}>0$ this equation has purely imaginary solutions
with respect to $\omega $. To avoid the vacuum instability, below we will
assume that $A_{s}\leq 0$. Under this condition all solutions to (\ref{modeq}%
) are real. The coefficient $C_{D}$ in Eq. (\ref{wavesracture}) is
determined by the normalization condition and is equal to \cite{Saha02}
\begin{equation}
C_{D}^{2}=\frac{1}{(2\pi )^{D-1}}\frac{\bar{I}_{i\omega _{n}}(ka)}{\frac{%
\partial }{\partial \omega }\bar{K}_{i\omega }(ka)|_{\omega =\omega _{n}}},
\label{normcoef}
\end{equation}%
where for a given function $F(x)$ we use the notation
\begin{equation}
\bar{F}(x)=AF(x)+xF^{\prime }(x).  \label{barnot}
\end{equation}

Substituting the eigenfunctions into the mode-sum formula (\ref{modesumform}%
) and using the relation
\begin{equation}
K_{i\omega _{n}}(ka)\bar{I}_{i\omega _{n}}(ka)=1,  \label{rel1}
\end{equation}%
the vacuum expectation value of the surface energy-momentum tensor can be
presented in the form
\begin{equation}
\langle 0|\tau _{l}^{k}|0\rangle =\frac{B_{D}I_{{\rm R}}(A)}{a^{D}}\left[
2\zeta \delta _{l}^{0}\delta _{0}^{k}+(4\zeta -1)A\delta _{l}^{k}\right]
,\quad l,k=0,2,\ldots ,D,  \label{Tsurf}
\end{equation}%
with
\begin{equation}
I_{{\rm R}}(A)=\int_{0}^{\infty }dx\,x^{D-2}\sum_{n=1}^{\infty }\frac{%
K_{i\omega _{n}}(x)}{\frac{\partial }{\partial \omega }\bar{K}_{i\omega
}(x)|_{\omega =\omega _{n}}},  \label{IA1}
\end{equation}%
and $\langle 0|\tau _{1}^{1}|0\rangle =0$. Here and below the quantities for
the RR and RL regions are denoted by the indices R and L, respectively, and
we use the notation
\begin{equation}
B_{D}=\frac{1}{(4\pi )^{\frac{D-1}{2}}\Gamma \left( \frac{D-1}{2}\right) }.
\label{BD}
\end{equation}%
The surface energy-momentum tensor (\ref{Tsurf}) has a diagonal structure:
\begin{equation}
\langle 0|\tau _{l}^{k}|0\rangle ={\rm diag}\left( \varepsilon ,0,-p,\ldots
,-p\right) ,  \label{taudiag}
\end{equation}%
with the surface energy density $\varepsilon $ and stress $p$, and the
equation of state%
\begin{equation}
\varepsilon =-\left[ 1+\frac{2\zeta }{A(4\zeta -1)}\right] p.
\label{eqstate}
\end{equation}%
For a minimally coupled scalar field, this corresponds to a cosmological
constant induced on the plate. In accordance with (\ref{Tsurf}) for the
vacuum stress one has%
\begin{equation}
p=\frac{B_{D}}{a^{D}}A(1-4\zeta )I_{R}(A).  \label{surfstress}
\end{equation}

The quantity (\ref{IA1}) and, hence, the surface energy-momentum tensor
diverges and needs some regularization. Many regularization techniques are
available nowadays and, depending on the specific physical problem under
consideration, one of them may be more suitable than the others. In
particular, the generalized zeta function method \cite{Dowk76,Eliz94,More97}
is in general very powerful to give physical meaning to the divergent
quantities. There are several examples of the application of this method to
the evaluation of the Casimir effect (see, for instance, \cite%
{Eliz94,Blau88,Eliz93,Lese94,Bord96,Lese96,Bord96b,Bord97,
Lamb99,Cogn01}). Here we will use the method which is an analog of
the generalized zeta function approach. We define the function
\begin{equation}
F_{{\rm R}}(s)=\int_{0}^{\infty }dx\,x^{D-2}\zeta _{{\rm R}}(s,x),
\label{IAs}
\end{equation}%
where
\begin{equation}
\zeta _{{\rm R}}(s,x)=\sum_{n=1}^{\infty }\frac{\omega _{n}^{-s}K_{i\omega
_{n}}(x)}{\frac{\partial }{\partial \omega }\bar{K}_{i\omega }(x)|_{\omega
=\omega _{n}}}.  \label{zetsx}
\end{equation}%
The computation of vacuum expectation value for the surface energy-momentum
tensor requires an analytical continuation of the function $F_{{\rm R}}(s)$
to the value $s=0$,
\begin{equation}
I_{{\rm R}}(A)=F_{{\rm R}}(s)|_{s=0}.  \label{IFs0}
\end{equation}

The starting point of our consideration is the representation of the
function (\ref{zetsx}) in terms of contour integral
\begin{equation}
\zeta _{{\rm R}}(s,x)=\frac{1}{2\pi i}\int_{C}dz\,z^{-s}\frac{K_{iz}(x)}{%
\bar{K}_{iz}(x)},  \label{intzetsx1}
\end{equation}%
where $C$ is a closed counterclockwise contour in the complex $z$ plane
enclosing all zeros $\omega _{n}(x)$. The location of these zeros \ enables
one to deform the contour $C$ into a segment of the imaginary axis $(-iR,iR)$
and a semicircle of radius $R$ in the right half-plane. We will also assume
that the origin is avoided by the semicircle $C_{\rho }$ with small radius $%
\rho $. For sufficiently large $s$ the integral over the large semicircle in
(\ref{intzetsx1}) tends to zero in the limit $R\rightarrow \infty $, and the
expression on the right can be transformed to
\begin{equation}
\zeta _{{\rm R}}(s,x)=\frac{1}{2\pi i}\int_{C_{\rho }}dz\,z^{-s}\frac{%
K_{iz}(x)}{\bar{K}_{iz}(x)}-\frac{1}{\pi }\cos \frac{\pi s}{2}\int_{\rho
}^{\infty }dz\,z^{-s}\frac{K_{z}(x)}{\bar{K}_{z}(x)}.  \label{intzetsx2}
\end{equation}%
Below we will consider the limit $\rho \rightarrow 0$. In this limit the
first integral vanishes in the case $s=0$, and in the following we will
concentrate on the contribution of the second integral. For the analytic
continuation of this integral we employ the uniform asymptotic expansion of
the MacDonald function for large values of the order \cite{Abramowitz}. We
will rewrite this expansion in the form
\begin{equation}
K_{z}(x)\sim \sqrt{\frac{\pi }{2}}\frac{e^{-z\eta (x/z)}}{(x^{2}+z^{2})^{1/4}%
}\sum_{l=0}^{\infty }\frac{(-1)^{l}\widetilde{u}_{l}(t)}{(x^{2}+z^{2})^{l/2}}%
,  \label{Kzxasymp}
\end{equation}%
where
\begin{equation}
t=\frac{z}{\sqrt{x^{2}+z^{2}}},\quad \eta (x)=\sqrt{1+x^{2}}+\ln \frac{x}{1+%
\sqrt{1+x^{2}}},\quad \tilde{u}_{l}(t)=\frac{u_{l}(t)}{t^{l}},
\label{tetaul}
\end{equation}%
and the expressions for the functions $u_{l}(t)$ are given in \cite%
{Abramowitz}. From these expressions it follows that the coefficients $%
\tilde{u}_{l}(t)$ have the structure
\begin{equation}
\tilde{u}_{l}(t)=\sum_{m=0}^{l}u_{lm}t^{2m},  \label{ult}
\end{equation}%
with numerical coefficients $u_{lm}$. From Eq. (\ref{Kzxasymp}) and the
corresponding expansion for the derivative of the MacDonald function we
obtain the asymptotic expansion
\begin{equation}
\bar{K}_{z}(x)\sim -\sqrt{\frac{\pi }{2}}(x^{2}+z^{2})^{1/4}e^{-z\eta
(x/z)}\sum_{l=0}^{\infty }\frac{(-1)^{l}\tilde{v}_{l}(t)}{(x^{2}+z^{2})^{l/2}%
}\,,  \label{Kzxbaras}
\end{equation}%
where
\begin{equation}
\tilde{v}_{l}(t)=\frac{v_{l}(t)}{t^{l}}+A\tilde{u}_{l-1}\,,  \label{vltbar}
\end{equation}%
and the expressions for $v_{l}(t)=t^{l}\sum_{m=0}^{l}v_{lm}t^{2m}$ are
presented in \cite{Abramowitz}. The recurrence formulae for the numerical
coefficients $u_{lm}$ and $v_{lm}$ can be found in Ref. \cite{Saha04a}. Note
that the functions (\ref{vltbar}) have the structure
\begin{equation}
\tilde{v}_{l}(t)=\sum_{m=0}^{l}\tilde{v}_{lm}t^{2m},\quad \tilde{v}%
_{lm}=v_{lm}+Au_{l-1,m}\,.  \label{vltbar1}
\end{equation}%
From Eqs. (\ref{Kzxasymp}) and (\ref{Kzxbaras}) we can find the asymptotic
expansion for the ratio in the second integral on the right of formula (\ref%
{intzetsx2}). For the further convenience we will write this expansion in
the form
\begin{equation}
\frac{K_{z}(x)}{\bar{K}_{z}(x)}\sim -\frac{1}{(x^{2}+z^{2})^{1/2}}%
\sum_{l=0}^{\infty }\frac{(-1)^{l}U_{l}(t)}{(1+x^{2}+z^{2})^{l/2}}\,,
\label{Kzratio}
\end{equation}%
where the coefficients $U_{l}(t)$ are defined by the relation
\begin{equation}
\sum_{l=0}^{\infty }(-1)^{l}\frac{\tilde{u}_{l}(t)}{r^{l}}\left[
\sum_{l=0}^{\infty }(-1)^{l}\frac{\tilde{v}_{l}(t)}{r^{l}}\right]
^{-1}=\sum_{l=0}^{\infty }\frac{(-1)^{l}U_{l}(t)}{(1+r^{2})^{l/2}}\,,
\label{defUl}
\end{equation}%
and similar to (\ref{ult}), (\ref{vltbar1}), are polynomials in $t$:%
\begin{equation}
U_{l}(t)=\sum_{m=0}^{l}U_{lm}t^{2m}.  \label{Ulmdef}
\end{equation}%
The first three coefficients are given by expressions
\begin{eqnarray}
U_{1}(t) &=&\frac{1}{2}-A-\frac{t^{2}}{2}\,,  \nonumber  \label{Ufunc} \\
U_{2}(t) &=&\frac{3}{8}-A+A^{2}+\left( -\frac{5}{4}+A\right) t^{2}+\frac{%
7t^{4}}{8}\,, \\
U_{3}(t) &=&\frac{5}{8}-\frac{3A}{2}+\frac{3A^{2}}{2}-A^{3}+\left( -\frac{25%
}{8}+3A-\frac{3A^{2}}{2}\right) t^{2}+\left( \frac{41}{8}-2A\right) t^{4}-%
\frac{21t^{6}}{8}\,.  \nonumber
\end{eqnarray}

Now let us consider the function
\begin{equation}
F_{{\rm R}}(s)=-\frac{1}{\pi }\cos \frac{\pi s}{2}\int_{0}^{\infty
}dx\,x^{D-2}\int_{\rho }^{\infty }dz\,z^{-s}\frac{K_{z}(x)}{\bar{K}_{z}(x)}.
\label{Fs}
\end{equation}%
We subtract and add to the integrand in this equation the corresponding
asymptotic expression. This allows us to split (\ref{Fs}) into the following
pieces
\begin{equation}
F_{{\rm R}}(s)=F_{{\rm R}}^{(as)}(s)+F_{{\rm R}}^{(1)}(s)\,,  \label{FasF1}
\end{equation}%
where
\begin{eqnarray}
F_{{\rm R}}^{(as)}(s) &=&\frac{1}{\pi }\cos \frac{\pi s}{2}\int_{0}^{\infty
}dx\,x^{D-2}\int_{\rho }^{\infty }dz\,z^{-s}\frac{1}{r}\sum_{l=0}^{N}\frac{%
(-1)^{l}U_{l}(\cos \theta )}{(1+r^{2})^{l/2}},  \label{Fas} \\
F_{{\rm R}}^{(1)}(s) &=&-\frac{1}{\pi }\cos \frac{\pi s}{2}\int_{0}^{\infty
}dx\,x^{D-2}\int_{\rho }^{\infty }dz\,z^{-s}\left[ \frac{K_{z}(x)}{\bar{K}%
_{z}(x)}+\frac{1}{r}\sum_{l=0}^{N}\frac{(-1)^{l}U_{l}(\cos \theta )}{%
(1+r^{2})^{l/2}}\right] ,  \label{F1}
\end{eqnarray}%
and
\begin{equation}
r=(x^{2}+z^{2})^{1/2},\quad \cos \theta =z/r.  \label{rtet}
\end{equation}%
For $N\geq D-1$ the expression for $F_{{\rm R}}^{(1)}(s)$ is finite at $s=0$
and, hence, for our aim it is sufficient to subtract $N=D-1$ asymptotic
terms. At $s=0$ the function $F_{{\rm R}}^{(1)}(s)$ is finite for $\rho =0$
and we can directly put this value. The integral over $z$ in the expression
for $F_{{\rm R}}^{(as)}(s)$ is finite in the limit $\rho \rightarrow 0$ for $%
0<{\rm Re\,}s<1$. For these values we can put $\rho =0$ in Eq. (\ref{Fas}).
After the integration over $z$, the contribution of the $l=0$ term in this
formula can be presented in the form%
\begin{equation}
\frac{1}{2\pi }B\left( \frac{1-s}{2},\frac{s}{2}\right) \cos \frac{\pi s}{2}%
\int_{0}^{\infty }dx\,x^{D-2-s},  \label{l0asterm}
\end{equation}%
with the beta function $B(x,y)$. Now using the standard dimensional
regularization result that the renormalized value of the integrals of the
type $\int_{0}^{\infty }dx\,x^{\beta }$ is equal to zero (see, e.g., \cite%
{Coll84}), we conclude that the contribution of the term with $l=0$ into Eq.
(\ref{Fas}) vanishes. This can be seen by another way, considering the case
of a scalar field with nonzero mass $m$ and taking the limit $m\rightarrow 0$
after the evaluation of the corresponding integrals (for this trick in the
calculations of the Casimir energy see, for instance, Refs. \cite%
{Bord02,Nest03}). For the massive case, in Eq. (\ref{wavesracture}) one has $%
\phi (\xi )=K_{i\omega }(\xi \sqrt{k^{2}+m^{2}})$ and the corresponding
formulae are obtained from those given above in this section by replacement $%
x\rightarrow \sqrt{x^{2}+m^{2}a^{2}}$. With this replacement the
integral corresponding to the contribution of $l=0$ term into Eq. (\ref{Fas}%
) can be easily evaluated in terms of the gamma function and vanishes in the
limit $m\rightarrow 0$ for ${\rm Re}\,s<D-1$. After performing to the polar
coordinates $(r,\theta )$ and using relation (\ref{Ulmdef}), from (\ref{Fas}%
) one finds
\begin{equation}
F_{{\rm R}}^{(as)}(s)=\frac{1}{\pi }\cos \frac{\pi s}{2}%
\sum_{l=1}^{N}(-1)^{l}\sum_{m=0}^{l}U_{lm}\int_{0}^{\infty }dr\frac{r^{D-s-2}%
}{(1+r^{2})^{l/2}}\int_{0}^{\pi /2}d\theta \sin ^{D-2}\theta \cos
^{2m-s}\theta .  \label{Fas1}
\end{equation}%
Evaluating the integrals by using the standard formulae, we find the
expression
\begin{equation}
F_{{\rm R}}^{(as)}(s)=\frac{1}{\pi }\cos \frac{\pi s}{2}%
\sum_{l=1}^{N}(-1)^{l}B\left( \frac{D-s-1}{2},\frac{l+s-D+1}{2}\right)
\sum_{m=0}^{l}U_{lm}B\left( m-\frac{s-1}{2},\frac{D-1}{2}\right) ,
\label{Fas2}
\end{equation}%
where the pole contribution is given explicitly in terms of the beta
function. In the sum over $l$, the terms with odd $D-l\geq 1$ have simple
poles at $s=0$ coming from the first beta function. Introducing a new
summation variable $j=(D-l-1)/2$, the corresponding residue can be easily
found by using the standard formula for the gamma function:
\begin{equation}
F_{{\rm R},-1}^{(as)}=-\frac{2}{\pi }(-1)^{D}\Gamma \left( \frac{D-1}{2}%
\right) \sum_{j=0}^{j_{D}}\frac{(-1)^{j}}{\Gamma (j+1)\Gamma \left( \frac{D-1%
}{2}-j\right) }\sum_{m=0}^{D-2j-1}U_{D-2j-1,m}B\left( m+\frac{1}{2},\frac{D-1%
}{2}\right) ,  \label{Fas-1}
\end{equation}%
where
\begin{equation}
j_{D}=\left\{
\begin{array}{ll}
\frac{D-2}{2}, & {\rm for\,even}\quad D \\
\frac{D-3}{2}, & {\rm for\,odd}\quad D\
\end{array}%
\right. .  \label{pD}
\end{equation}%
Laurent-expanding near $s=0$ we can write
\begin{equation}
F_{{\rm R}}^{(as)}(s)=\frac{F_{{\rm R},-1}^{(as)}}{s}+F_{{\rm R},0}^{(as)}+%
{\cal {O}}(s),  \label{Fas3}
\end{equation}%
with
\begin{eqnarray}
F_{{\rm R},0}^{(as)} &=&-\frac{(-1)^{D}}{\pi }\Gamma \left( \frac{D-1}{2}%
\right) \sum_{j=0}^{j_{D}}\frac{(-1)^{j}}{\Gamma (j+1)\Gamma \left( \frac{D-1%
}{2}-j\right) }\sum_{m=0}^{D-2j-1}U_{D-2j-1,m}B\left( m+\frac{1}{2},\frac{D-1%
}{2}\right)   \nonumber \\
&\times &\left[ \psi \left( m+\frac{D}{2}\right) +\psi \left( j+1\right)
-\psi \left( m+\frac{1}{2}\right) -\psi \left( \frac{D-1}{2}\right) \right]
\label{Fas0} \\
&+&\frac{1}{\pi }\left( \sum_{l=1,D-l={\rm even}}^{D-1}+\sum_{l=D}^{N}%
\right) (-1)^{l}B\left( \frac{l-D+1}{2},\frac{D-1}{2}\right)
\sum_{m=0}^{l}U_{lm}B\left( m+\frac{1}{2},\frac{D-1}{2}\right) ,  \nonumber
\end{eqnarray}%
where $\psi (x)=d\ln \Gamma (x)/dx$ is the diagamma function and the second
sum in the braces of the third line is present only for $N\geq D$. The first
term on the right of Eq. (\ref{Fas0}) with diagamma functions comes from the
finite part of the Laurent expansion of the summands with odd $D-l$ in Eq. (%
\ref{Fas2}). Gathering all contributions together, near $s=0$ we find
\begin{equation}
F_{{\rm R}}(s)=\frac{F_{{\rm R},-1}^{(as)}}{s}+F_{{\rm R},0}^{(as)}+F_{{\rm R%
}}^{(1)}(0)+{\cal {O}}(s).  \label{Fas4}
\end{equation}%
Using this result, for the vacuum stress induced on the surface of a single
plate one obtains%
\begin{equation}
p=p_{p}^{{\rm (R)}}+p_{f}^{{\rm (R)}},  \label{ppf}
\end{equation}%
where for the pole and finite contributions one has%
\begin{eqnarray}
p_{p}^{{\rm (R)}} &=&\frac{B_{D}}{sa^{D}}A(1-4\zeta )F_{{\rm R}%
,-1}^{(as)},\quad A=aA_{s},  \label{ppf1} \\
p_{f}^{{\rm (R)}} &=&\frac{B_{D}}{a^{D}}A(1-4\zeta )\left[ F_{{\rm R}%
,0}^{(as)}+F_{{\rm R}}^{(1)}(0)\right] ,
\end{eqnarray}%
and the coefficients are defined by expressions (\ref{F1}), (\ref{Fas-1}), (%
\ref{Fas0}). The corresponding formulae for the pole and finite parts of the
surface energy density are obtained by using the equation of state (\ref%
{eqstate}). In particular, in the case of the Neumann boundary condition ($%
A=0$) for the finite parts one has%
\begin{equation}
\varepsilon _{{\rm N}f}^{{\rm (R)}}=\frac{2\zeta B_{D}}{a^{D}}\left[ F_{{\rm %
R},0}^{(as)}+F_{{\rm R}}^{(1)}(0)\right] _{A=0},\quad p_{{\rm N}f}^{{\rm (R)}%
}=0.  \label{RRNeu}
\end{equation}

\begin{figure}[tbph]
\begin{center}
\epsfig{figure=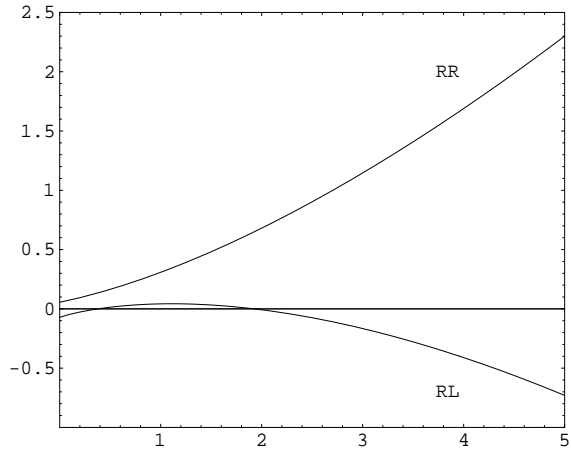,width=7.5cm,height=6cm}
\end{center}
\caption{Finite parts of the surface vacuum stresses $a^{D}p_{f}^{{\rm {(R,L)%
}}}/(A(1-4\protect\zeta ))$ for the RR and RL regions as functions
on $-aA_{s}$.} \label{figsurfen}
\end{figure}

In Fig. \ref{figsurfen} we have plotted the dependence of the finite part of
the quantity $a^{D}p_{f}^{{\rm (R)}}/[A(1-4\zeta )]$ on the parameter $%
-aA_{s}$ in the spatial dimension $D=3$. This quantity does not depend on
the curvature coupling parameter and is positive for the RR region. For a
minimally coupled scalar and $A>0$ this corresponds to the generation of the
negative cosmological constant on the plate. The corresponding numerical
value is determined by the proper acceleration of the plate. The surface
energy per unit surface of the plate can be found integrating the energy
density from Eq. (\ref{modesumform}),%
\begin{equation}
E^{{\rm (R,surf)}}=\int d^{D}x\sqrt{|\det g_{ik}|}\langle 0|T_{0}^{{\rm %
(surf)}0}|0\rangle =a\langle 0|\tau _{0}^{0}|0\rangle =a\varepsilon .
\label{ERsurf}
\end{equation}%
In the case of the Neumann boundary condition, for the finite part of the
surface energy in $D=3$ one obtains%
\begin{equation}
E_{{\rm N}f}^{{\rm (R,surf)}}=\frac{0.111\zeta }{a^{2}}.  \label{ERsurfN}
\end{equation}%
The finite part of the corresponding total vacuum energy per unit surface is
evaluated in Ref. \cite{Saha04a} and is equal to $E_{{\rm N}f}^{{\rm (R)}%
}=0.00213/a^{2}$. This quantity is the sum of the volume and surface
energies.

\section{Surface energy density and stresses in the RL region}

\label{sec:RLreg}

Consider the scalar vacuum in the region between the plate and the Rindler
horizon corresponding to $\xi =0$ (RL region). As in the previous section we
will assume that the field obeys boundary condition (\ref{boundcond}) on the
surface $\xi =a$. Note that for the RL region one has $n^{l}=-\delta _{1}^{l}
$, $K_{00}=-\xi $. To deal with discrete spectrum, we can introduce the
second plate located at $\xi =b<a$, on whose surface we impose boundary
conditions as well. After the construction of the corresponding partial zeta
function we take the limit $b\rightarrow 0$. As a result we can see that the
surface energy-momentum tensor in the RL region has a structure given by (%
\ref{taudiag}) and with the equation of state (\ref{eqstate}). For the
surface vacuum stress one obtains the expression
\begin{equation}
p=\frac{B_{D}}{a^{D}}A(1-4\zeta )I_{{\rm L}}(A),\quad A=-aA_{s},
\label{surfstressL}
\end{equation}%
where now
\begin{equation}
I_{{\rm L}}(A)=F_{{\rm L}}(s)|_{s=0},\quad F_{{\rm L}}(s)=\int_{0}^{\infty
}dx\,x^{D-2}\zeta _{{\rm L}}(s,x),  \label{IRL}
\end{equation}%
with
\begin{equation}
\zeta _{{\rm L}}(s,x)=-\frac{1}{\pi }\cos \frac{\pi s}{2}\int_{\rho
}^{\infty }dz\,z^{-s}\frac{I_{z}(x)}{\bar{I}_{z}(x)},  \label{zetaRL}
\end{equation}%
and the barred notation is in accordance with Eq. (\ref{barnot}). For a
given $A$ this expression differs from the corresponding expression for the
RR region by the replacement $K_{z}(x)\rightarrow I_{z}(x)$. Note that as in
the previous section, to avoid the vacuum instability, here we have assumed $%
A_{s}\leq 0$. Under this condition, for a given $x$ the function $\bar{I}%
_{z}(x)$ has no real positive zeros with respect to $z$. By using the
uniform asymptotic expansions for the Bessel modified function $I_{z}(x)$
and its derivative with respect to the argument, for the ratio in the
subintegrand of expression (\ref{zetaRL}) one receives
\begin{equation}
\frac{I_{z}(x)}{\bar{I}_{z}(x)}\sim \frac{1}{(x^{2}+z^{2})^{1/2}}%
\sum_{l=0}^{\infty }\frac{U_{l}(t)}{(1+x^{2}+z^{2})^{l/2}},  \label{Izratio}
\end{equation}%
with the same coefficients $U_{l}(t)$ as in Eq. (\ref{Kzratio}). Now the
expression for $F_{{\rm L}}(s)$ can be written as
\begin{equation}
F_{{\rm L}}(s)=F_{{\rm L}}^{(as)}(s)+F_{{\rm L}}^{(1)}(s),  \label{FasF1L}
\end{equation}%
where
\begin{eqnarray}
F_{{\rm L}}^{(as)}(s) &=&-\frac{1}{\pi }\cos \frac{\pi s}{2}\int_{0}^{\infty
}dx\,x^{D-2}\int_{\rho }^{\infty }dz\,z^{-s}\frac{1}{r}\sum_{l=0}^{N}\frac{%
U_{l}(\cos \theta )}{(1+r^{2})^{l/2}},  \label{FasL} \\
F_{{\rm L}}^{(1)}(s) &=&-\frac{1}{\pi }\cos \frac{\pi s}{2}\int_{0}^{\infty
}dx\,x^{D-2}\int_{\rho }^{\infty }dz\,z^{-s}\left[ \frac{I_{z}(x)}{\overline{%
I}_{z}(x)}-\frac{1}{r}\sum_{l=0}^{N}\frac{U_{l}(\cos \theta )}{%
(1+r^{2})^{l/2}}\right] .  \label{F1L}
\end{eqnarray}%
By the way similar to the case of the RR region, the asymptotic part can be
presented in the form
\begin{equation}
F_{{\rm L}}^{(as)}(s)=-\frac{1}{\pi }\cos \frac{\pi s}{2}\sum_{l=1}^{N}B%
\left( \frac{D-s-1}{2},\frac{l+s-D+1}{2}\right) \sum_{m=0}^{l}U_{lm}B\left(
m-\frac{s-1}{2},\frac{D-1}{2}\right) .  \label{Fas2L}
\end{equation}%
The corresponding residue is easily found by using the formula for the gamma
function:
\begin{equation}
F_{{\rm L},-1}^{(as)}=-\frac{2}{\pi }\Gamma \left( \frac{D-1}{2}\right)
\sum_{j=0}^{j_{D}}\frac{(-1)^{j}}{\Gamma (j+1)\Gamma \left( \frac{D-1}{2}%
-j\right) }\sum_{m=0}^{D-2j-1}U_{D-2j-1,m}B\left( m+\frac{1}{2},\frac{D-1}{2}%
\right) ,  \label{Fas-1L}
\end{equation}%
with $j_{D}$ defined by expression (\ref{pD}). Expanding near the simple
pole $s=0$ we can write
\begin{equation}
F_{{\rm L}}^{(as)}(s)=\frac{F_{{\rm L},-1}^{(as)}}{s}+F_{{\rm L},0}^{(as)}+%
{\cal {O}}(s),  \label{Fas3L}
\end{equation}%
with
\begin{eqnarray}
F_{{\rm L},0}^{(as)} &=&-\frac{1}{\pi }\Gamma \left( \frac{D-1}{2}\right)
\sum_{j=0}^{j_{D}}\frac{(-1)^{j}}{\Gamma (j+1)\Gamma \left( \frac{D-1}{2}%
-j\right) }\sum_{m=0}^{D-2j-1}U_{D-2j-1,m}B\left( m+\frac{1}{2},\frac{D-1}{2}%
\right)   \nonumber \\
&\times &\left[ \psi \left( m+\frac{D}{2}\right) +\psi \left( j+1\right)
-\psi \left( m+\frac{1}{2}\right) -\psi \left( \frac{D-1}{2}\right) \right]
\label{Fas0L} \\
&-&\frac{1}{\pi }\left( \sum_{l=1,D-l={\rm even}}^{D-1}+\sum_{l=D}^{N}%
\right) B\left( \frac{l-D+1}{2},\frac{D-1}{2}\right)
\sum_{m=0}^{l}U_{lm}B\left( m+\frac{1}{2},\frac{D-1}{2}\right) ,  \nonumber
\end{eqnarray}%
where the second sum in the braces of the third line is present only for $%
N\geq D$. Taking all contributions together, near $s=0$ we find
\begin{equation}
F_{{\rm L}}(s)=\frac{F_{{\rm L},-1}^{(as)}}{s}+F_{{\rm L},0}^{(as)}+F_{{\rm L%
}}^{(1)}(0)+{\cal {O}}(s),  \label{Fas4L}
\end{equation}%
with different terms defined by formulae (\ref{F1L}), (\ref{Fas-1L}), (\ref%
{Fas0L}). Note that for a given $A$ the pole parts of the surface densities
for the RR and RL regions differ by the factor $(-1)^{D}$. Relation (\ref%
{Fas4L}) allows us to present the vacuum stress as a sum of pole and finite
parts:
\begin{eqnarray}
p_{p}^{{\rm (L)}} &=&\frac{B_{D}}{sa^{D}}A(1-4\zeta )F_{{\rm L}%
,-1}^{(as)},\quad A=-aA_{s}  \label{ppfL} \\
p_{f}^{{\rm (L)}} &=&\frac{B_{D}}{a^{D}}A(1-4\zeta )\left[ F_{{\rm L}%
,0}^{(as)}+F_{{\rm L}}^{(1)}(0)\right] .
\end{eqnarray}%
The corresponding decomposition of the surface energy density is obtained
from here with the help of the equation of state (\ref{eqstate}). In Fig. %
\ref{figsurfen}, the finite part of $D=3$ surface vacuum stress $a^{D}p_{f}^{%
{\rm (L)}}/[A(1-4\zeta )]$ is presented for the RL region as a
function of the parameter $-aA_{s}$. In the case of a minimally
coupled scalar the corresponding surface energy-momentum tensor is
a cosmological constant type located on the plate. As follows from
Fig. \ref{figsurfen}, in $D=3$ the induced cosmological constant
is negative for $0.378<A<1.941$ and is positive otherwise. For the
Neumann boundary condition one has a relation similar to
(\ref{RRNeu}) with the replacement ${\rm R}\rightarrow {\rm L}$.
In this case the finite part of the surface energy per unit area
of the plate for $D=3$ is given by formula%
\begin{equation}
E_{{\rm N}f}^{{\rm (L,surf)}}=-\frac{0.142\zeta }{a^{2}}.  \label{ELsurfN}
\end{equation}%
In accordance with Ref. \cite{Saha04a} the corresponding part in
the total vacuum energy is equal to $E_{{\rm N}f}^{{\rm
(L)}}=0.000792/a^{2}$. For an infinitely thin plate taking the RR
and RL regions together, the pole parts of the both surface and
total vacuum energies cancel in $D=3$ and the corresponding
Casimir energies are finite:%
\begin{equation}
E_{{\rm N}}^{{\rm (surf)}}=-\frac{0.0309\zeta }{a^{2}},\quad E_{{\rm N}}=%
\frac{0.00292}{a^{2}}.  \label{EsurfNRL}
\end{equation}%
The relation of these quantities to the energies measured by a uniformly
accelerated observer is discussed in Ref. \cite{Saha04a}.

\section{Conclusion}

\label{sec:Conc}

In this paper we have investigated the surface Casimir densities generated
by a single plate moving by uniform proper acceleration, assuming that the
field is prepared in the Fulling-Rindler vacuum state. The corresponding
volume vacuum expectation values of the energy--momentum tensor were
investigated in Refs. \cite{Candelas,Saha02} for the geometry of a single
plate and in Ref. \cite{Avag02} in the case of two plates. Here we consider
a scalar field with mixed boundary conditions and as a regularization method
the zeta function technique is employed. In the case of a single plate
geometry the right Rindler wedge is divided into two regions, referred as RR
and RL regions. By using the Cauchy's theorem on residues, we have
constructed an integral representations for the corresponding zeta functions
in both these regions, which are well suited for the analytic continuation.
Subtracting and adding to the integrands the leading terms of the
corresponding uniform asymptotic expansions, we present the corresponding
functions as a sum of two parts. The first one is convergent at the physical
point and can be evaluated numerically. In the second, asymptotic part the
pole contributions are given explicitly in terms of the beta function. As a
consequence, the vacuum expectation values of the surface energy-momentum
tensor for separate RR and RL regions contain pole and finite contributions.
The remained pole term is a characteristic feature for the zeta function
regularization method and has been found for many other cases of boundary
geometries. For a minimally coupled scalar field, the surface
energy-momentum tensor induced by quantum vacuum effects corresponds to a
source of a cosmological constant type located on the plate and with the
cosmological constant determined by the proper acceleration of the plate. In
the case of the Neumann boundary condition the finite parts of the surface
vacuum stresses vanish for the both RR and RL regions. In $D=3$ spatial
dimensions the corresponding surface energies are given by relations (\ref%
{ERsurfN}) and (\ref{ELsurfN}). In this case for an infinitely
thin plate taking the RR and RL regions together the pole parts of
the surface vacuum energies cancel and the total surface energy is
finite. The corresponding total vacuum energy, including the
surface and volume parts, is evaluated in Ref. \cite{Saha04a}. By
using the conformal relation between the Rindler and dS spacetimes
and the results from \cite{Saha02}, in Ref. \cite{Saha04c} the
vacuum expectation value of the bulk energy-momentum tensor is
evaluated for a conformally coupled scalar field which satisfies
the Robin boundary condition on a curved brane in dS spacetime. By
making use the same technique and the conformal properties of the
surface energy-momentum tensor, from the results of the present
paper we can obtain the surface vacuum energy-momentum tensor
induced on the brane in dS spacetime, which is a conformal image
of a uniformly accelerated plate in the Minkowski spacetime.

\section*{Acknowledgement}

The work of A.A.S. was supported in part by the Armenian Ministry of
Education and Science (Grant No.~0887).

\end{document}